\documentclass[12pt,a4paper]{article}
\usepackage[cp1251]{inputenc}
\usepackage{pazh}
\tightenlines
\usepackage{graphicx}
    \usepackage{epstopdf}
\usepackage{amssymb}
\usepackage{xcolor}

\tightenlines
\newcommand{\WI}[2]{#1_{\mathrm{#2}}}
\newcommand{\isn}[2]{\mbox{$^{#2}${#1}}}

\input epsf
\begin{document}
\baselineskip 21pt

\title{Production of Heavy Elements During the Explosion of a Low-Mass
Neutron Star in a Close Binary}

\author{\bf
I.V. Panov\affilmark{1,2*}, A.V. Yudin\affilmark{1,2}}

\affil{ {\it $^1$National Research Center ``Kurchatov Institute'' - ITEP}\\
{\it $^2$National Research Center ``Kurchatov Institute''}}

\vspace{2mm}

\vspace{2mm} \noindent
{The nucleosynthesis of heavy elements in the scenario for the evolution of a close binary of
neutron stars differing greatly in mass is considered. In contrast to the scenario for the merger of two
neutron stars of comparable masses considered repeatedly in the literature, the evolution of such a binary at
the final stage consists in a rapid mass transfer to the more massive star and an explosive disruption of the
low-mass component. We provide the details of the explosion and calculate the abundances of the heavy
elements produced in this process for various initial conditions.}



\vfill
\noindent\rule{8cm}{1pt}\\
{$^*$ email: panov\_iv@itep.ru}


\section*{INTRODUCTION}
The nucleosynthesis maintained by a rapid neutron
capture (the r-process) is res\-pon\-si\-ble for the production
of more than half of all elements heavier than
iron in nature. It results from the capture of neutrons
and the subsequent beta decays of forming short-lived
neutron-rich nuclei in an environment with a
high neutron number density. The region where it
proceeds on the map of nuclei lies near the neutron
stability boundary (Burbidge et al. 1957; Cameron
et al. 1957; Seeger et al. 1965).

A high initial neutron number density, up to 150
per seed nucleus (as a rule, such are the iron-peak
nuclei), is needed to create the conditions for the r-process
capable of producing heavy elements up to
the heaviest ones. Such conditions are achieved in
astrophysical objects in scenarios with a large neutron
excess and a high density of matter, for example,
during the merger of compact stellar remnants
in close binaries or during the explosions of
supernovae of a fairly rare type that form jets with a
high neutron number density (Thielemann et al. 2017;
Cowan et al. 2020) and in a hot wind from young
neutron stars (Cameron 2001; Arcones and Thielemann
2013).

The first detection of a neutron star (NS) merger
(Abbot et al. 2017) and the simultaneous observation
of r-elements (Tanvir et al. 2017) confirmed the understanding
that the main scenario for the r-process development is more likely associated with the ejecta
that form during a NS merger at the end of the
evolution of a close binary and not with supernova
explosions (Hudepohl et al. 2010).

The details of the NS merger scenario have long
been known (Freiburghaus et al. 1999) and the
conditions for the synthesis of heavy elements in
the r-process have been determined on their basis.
The NSs forming a close binary approach each other
due to the loss of angular momentum by the binary
through the radiation of gravitational waves. At the
final stage the stars merge into a single object --- a
supermassive NS or a black hole, with part of the
matter being ejected from the binary in the form of
jets or a wind. This is the so-called merging scenario
most popular at present.

However, there exists another possibility: if the
binary was initially highly asymmetric, then the stripping
scenario can be realized (see Clark and Eardley
1977). In this scenario the more massive and
compact star ``devours'' its less massive and more
extended companion. Losing its mass in the course
of such evolution, the latter approaches the innermost
stable configuration of a minimum-mass NS
and explodes. Blinnikov et al. (1984) first pointed to
the connection of this process with gamma-ray bursts
(GRBs).

In this paper we present the first nucleosynthesis
calculation in the stripping scenario. The paper is
structured as follows. In the next section we will
briefly describe the stripping scenario. Then, we will
consider the nucleosynthesis computation algorithm.
Next, we will describe the dynamics of the explosive
disruption of a low-mass NS. In the final section we
will present the results of our r-process calculations
and formulate our conclusions.

\section*{STRIPPING MODEL}
\noindent The final evolutionary stages of a NS binary attract
the increased attention of researchers. However,
in almost all of the multidimensional hydrodynamic
simulations performed so far the NS masses were
close and fairly large, $M\gtrsim M_{\odot}$, and the result of
their interaction was a merger (Korobkin et al. 2012;
Rosswog et al. 2014; Martin et al. 2015). Indeed,
the radius of such NSs depends weakly on the mass
(Lattimer and Prakash 2001) and they behave like
two liquid droplets when in contact, merging into a
single object—a supermassive NS or a black hole.

However, if the binary is highly asymmetric, i.e.,
the component masses differ sig\-ni\-fi\-can\-tly and, moreover,
the low-mass NS mass is sufficiently small,
then the stripping scenario can be realized (Clark and
Eardley 1977). As the binary components approach
each other, the lower-mass NS is the first to overfill its
Roche lobe and begins to flow onto the more massive
companion. During such a mass transfer it can reach
the lower NS mass limit (0.1~$M_\odot$; see, e.g. Haensel
et al. 2007) and explode, actually producing a GRB
(Blinnikov et al. 1984, 1990).

The interest in the stripping model was renewed
after the historic identification of the gravitational signal GW170817
and GRB170817A (Abbot et al. 2017).
Observations showed that many parameters of this
GRB, which turned out to be very peculiar, are
close to the predictions of the stripping model (for
a discussion, see Yudin et al. 2019).

The results of our hydrodynamic simulations of the
explosive disruption of a minimum-mass NS (see also
Blinnikov et al. 1990; Sumiyoshi et al. 1998) were
used to determine the pattern of nucleosynthesis in
this case.

\section*{NUCLEOSYNTHESIS COMPUTATION ALGORITHM}
\noindent Under conditions of a high neutron number density
the nucleosynthesis in the r-process proceeds
near the neutron stability boundary, which requires
predicting all characteristics of short-lived experimentally
unstudied nuclei. An additional complexity
in simulating the r-process in a highly neutron-rich
environment, which is also typical for a NS merger,
is fission. As was first shown in numerical simulations
of the r-process in the NS merger scenario
(see, e.g., Freiburghaus et al. 1999), the fission of heavy-element nuclei leads to nucleosynthesis cycling
(Panov et al. 2003), i.e., to the drawing of a
large number of fission product nuclei as new seed
nuclei into the r-process and to the production of
most heavy elements from the second peak in the
elemental abundance curve to thorium and uranium.
The corresponding increase of the theoretical data
used in the simulations, such as the rates of induced,
delayed, and spontaneous fission as well as
the mass distribution of fission product nuclei, and
their inclusion as new seed nuclei (Panov et al. 2003,
2008, 2009) makes the system of equations and the
simulation process more complicated and requires
optimizing the numerical schemes and algorithms.

For our numerical simulations of the r-process
we applied the nuclear reaction network previously
implemented in the SYNTHeZ code (Blinnikov and
Panov 1996; Nadyozhin et al. 1998), which specifies
the number densities of all the nuclei drawn into the
nuc\-le\-o\-syn\-the\-sis, including the neutron number density
control. In the updated SYNTHeR (nucleoSYNThesis
of HEavy elements in the R-process) code
(Korneev and Panov 2011) the fission reactions were
supplemented by a proper allowance for the mass
distribution of fission product nuclei and their return
to the r-process as new seed nuclei, which leads to
the establishment of a quasi-steady current of nuclei.

Since the nucleosynthesis is studied in the scenarios
of both explosive nucleosynthesis at high temperatures
and densities and the transition from the
explosive synthesis of elements to the alpha-process
and the r-process, the nucleosynthesis codes were
sup\-p\-le\-men\-ted by the reactions with charged particles
and the previously disregarded interaction of nucleons
and nuclei with electrons (Panov et al. 2018).
We supplemented the SYNTHeR code by the weak
interaction reactions, whose bank (Langanke and
Martinez-Pinedo 2000) contains data for the isotopes
of the iron-peak elements $(20<Z<32)$. There are
no lighter elements $(Z<20)$ in this bank, although,
as has been recently shown (Fischer et al. 2016), the
role of light elements is also quite significant and,
therefore, their reactions with electrons should also
be taken into account. The weak processes are especially
important at high temperatures ($T > 5 \times 10^9$~K)
and densities ($\rho > 10^8$~g/cm$^3$) and lead, in particular,
to a change in the electron fractions  $\WI{Y}{e}$.

Since the reaction rates of the listed processes
determining the eigenvalues of the Jacobi matrix for
the system of differential equations realized in our nucleosynthesis
codes differ in absolute value by orders
of magnitude, the system of nucleosynthesis equations
is a classic example of a stiff system of ordinary
differential equations (ODEs). Many methods have
been developed for the numerical integration of stiff
systems of ODEs; the method by Gear (1971) was
recognized as one of the most efficient ones.

The predictor–corrector method with the automatic
choice of a step and an order of the method
underlies the algorithm. The main difficulty in implementing
this algorithm is the necessity of solving a
very large system of linear equations (of the order of
several thousand—by the number of equations in the
nucleosynthesis system) when performing the corrector
iterations. Since the matrix of coefficients in
this system is sparse, special methods developed for
sparse matrices (see, e.g., Pissanetzky 1984), in particular,
the special software package for astrophysical
problems (Blinnikov et al. 1993), which allowed
some algorithms to be accelerated by 1–2 orders of
magnitude, were applied for its solution. Note that
the choice of a method for solving a sparse system
has a decisive influence on the efficiency of the entire
algorithm for the kinetic problem.

The popular methods (Gibbs et al. 1976) suggesting
that the matrix has a symmetric structure are
not always a good approximation for a real problem
(Lyutostanskii et al. 1986). Therefore, we chose
the algorithm for sparse matrices with an arbitrary
structure (Osterby and Zlatev 1983) implemented in
both SYNTHeZ and SYNTHeR codes that have an
internal check for the conservation

The boundaries of the region of the nuclides involved
in the nucleosynthesis were defined as $\WI{Z}{min} = 1$ and $\WI{Z}{max} = 110$, while $\WI{A}{min}$ and $\WI{A}{max}$ were specified
according to the mass model used: the extended
Thomas–Fermi model with the Strutinsky integral
(Aboussir et al. 1995) or the liquid-drop model
(Moeller et al. 1995). Thus, the total number of nuclei
N involved in the nucleosynthesis was determined.

The nuclear reaction rates, which are the coefficients
in the differential equations, were calculated
for the same mass models. All pair and other major
burning reactions, along with the alpha-decay and
fission reactions and weak interactions, enter into the
list of involved nuclear reactions. They include: all
pair reactions with neutrons, protons, alpha particles,
and gamma-ray photons; the beta-decay and beta-delayed
processes, such as the emission of several
neutrons during beta decay and delayed fission; the
induced and spontaneous fission; a number of other
important reactions, such as the 3--$\alpha$ reaction and the
\isn{C}{12}, \isn{O}{16} and \isn{Si}{28} burning reactions.

The applied scheme allows the nucleosynthesis to
be efficiently calculated in various scenarios at $T_9<7$
and $\rho < 10^{12}~$g/cm$^3$. Our main calculations were
made with the widely used rates of beta decay and delayed
neutrons (Moeller et al. 1997, 2003), alpha decay
(Moeller et al. 2003), the rates of thermonuclear
reactions (Rauscher and Thielemann 2000) and
fission (Panov et al. 2005, 2010, 2013; Korneev
and Panov 2011). The experimentally measured
beta decay rates were taken from the NuDat2 nuclear
database (2009). The neutron-capture rates by heavy
nuclei (for elements with $Z > 83$) and the neutron-induced
fission rates are based on the calculations by
Panov et al. (2010), while the delayed fission rates
were taken from Panov et al. (2005, 2010).

\section*{EXPLOSION OF A LOW-MASS NS}
\begin{figure}[ht]
	\begin{center}
		\includegraphics[width=.95\textwidth]{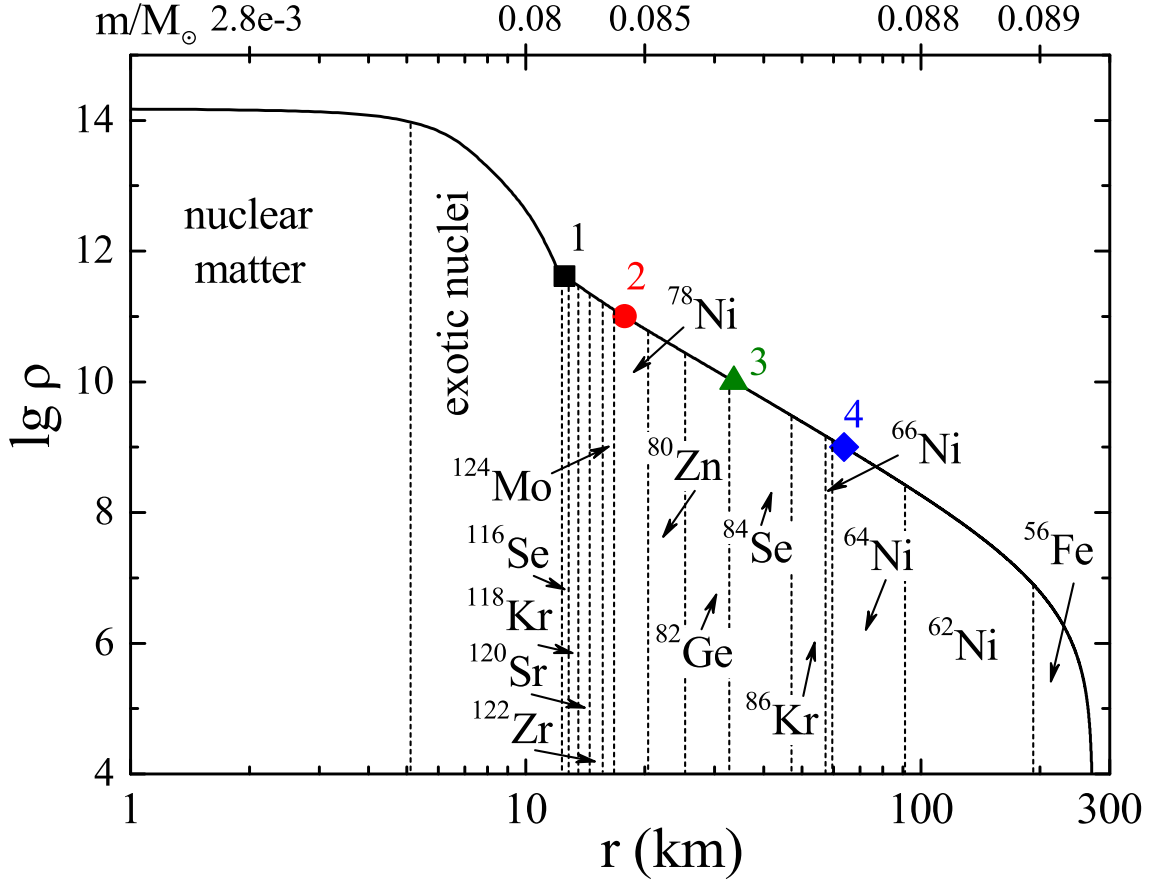}  %
	\end{center}
	\caption{Structure of a minimum-mass NS. The symbols mark the initial positions of the four trajectories used to calculate the
nucleosynthesis.} \label{Pix-LMNS}
\end{figure}

\noindent A minimum-mass NS has a peculiar structure.
Figure~\ref{Pix-LMNS} shows the dependence of the logarithm
of the density $\lg \rho$ on radial coordinate $r$ calculated
using the equation of state from Haensel and
Potekhin (2004). The corresponding Lagrangian
(mass) coordinates $m$ are shown on the upper axis.
The structure of the matter is clearly seen: the NS
core composed of nuclear matter lies at the center.
The core is surrounded by a mantle: a layer of exotic
nuclear configurations (``lasagne'', ``pasta'' etc.; see
Haensel et al. 2007) and hyper-nuclei immersed into a
sea of free neutrons. This part, containing 90\% 
of the
mass of the entire star, has a radius slightly greater
than 10~km.

The outer crust consists of a sequence of mono-layers
of nuclei (Ruster et al. 2006), from highly neutron-rich
isotopes, such as \isn{Se}{118}, to \isn{Fe}{56} on the surface.
The specific sequence of nuclei can slightly change,
depending on the mass formula used and other parameters
of the equation of state for the NS crust
(see, e.g., Martin et al. 2015), but the general trend
remains the same.

The symbols of different shapes in the figure indicate
the initial data for the four trajectories used to
calculate the nucleosynthesis. Some of their parameters
are given in Table 1.
\begin{table}[ht]
\centering
{{\bf Table 1.} Parameters of the trajectories used to calculate
the nucleosynthesis}
\vspace{5mm}\begin{tabular}{|c|c|c|c|c|c|} \hline\hline
Variant, & Initial & $T_9^\mathrm{max}$ & $\rho_0^{\rm max}$, & $r_0$, & $\WI{Y}{e}$\\
 No & composition &   & g/cm$^3$ &   km  &  \\
\hline
 1   &   \isn{Se}{116}     &   0.93   &   $4\times 10^{11}$ &  12.5   &  0.25  \\
\hline
 2   &   \isn{Ni}{78}     &   2.5    &   $10^{11}$     &   17.8  & 0.335 \\
\hline
 3   &    \isn{Se}{84}    &  6.3  &      $10^{10}$   &  33.8  & 0.405 \\
\hline
 4   &   \isn{Ni}{64}   &    10         &    $10^{9}$  &  63.5  &  0.44  \\
\hline
\end{tabular}
\label{Table_Trajectories}
\end{table}

\begin{figure}[ht]
	\begin{center}
	\includegraphics[width=.95\textwidth]{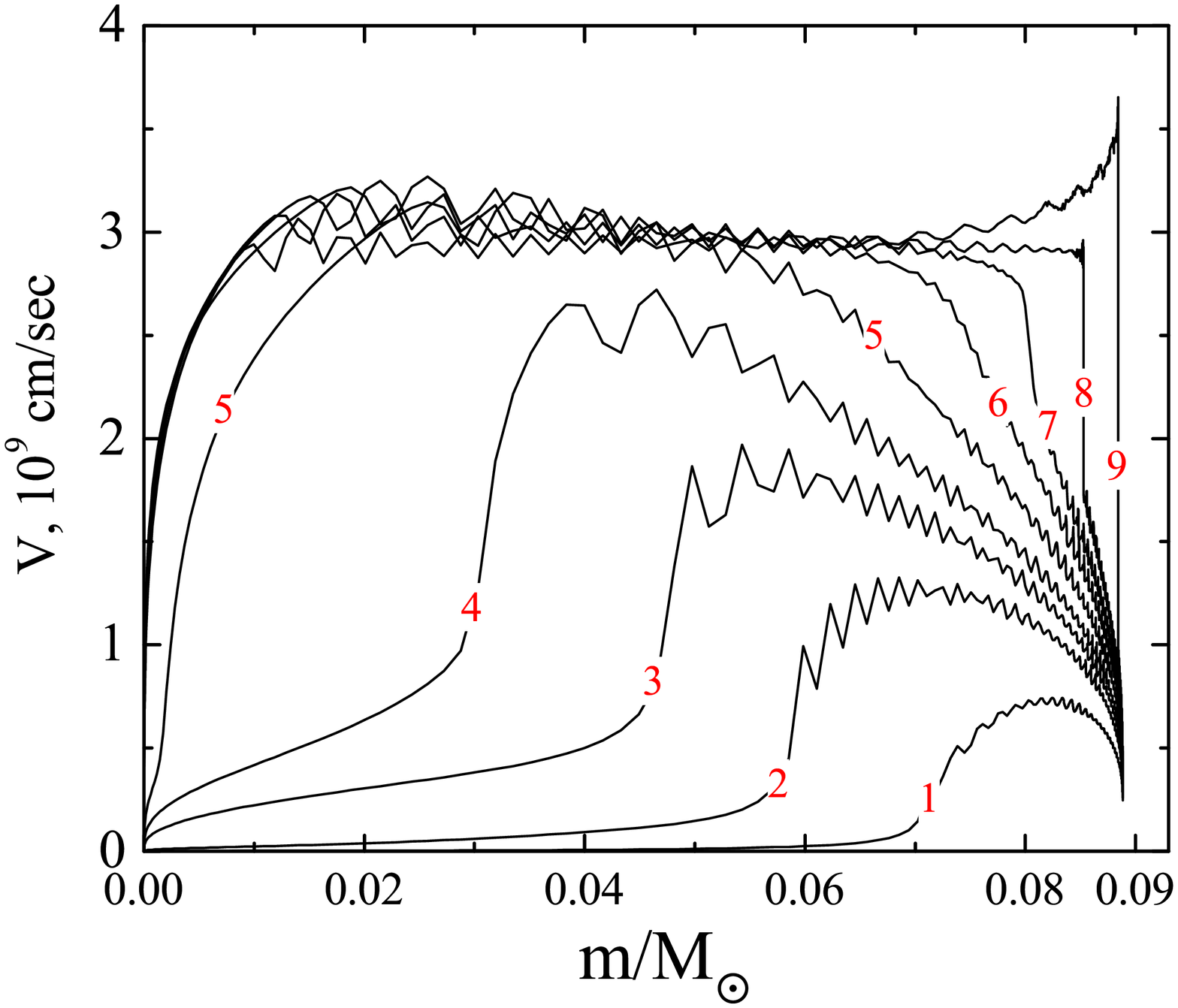}
	\end{center}
	\caption{Expansion velocity profiles for the NS matter as a function of mass coordinate $m$. The numbers mark the times (in
seconds): $t_1=0.360$, $t_2=0.365$, $t_3=0.368$, $t_4=0.369$,
$t_5=0.371$, $t_6=0.373$, $t_7=0.375$, $t_8=0.376$,  $t_9=0.378$.} \label{Pix-Vel}
\end{figure}

Consider the explosive disruption of a minimum mass
NS by assuming the problem to be spherically
symmetric (see Yudin et al. 2019). The dynamics
of this process was first computed by Blinnikov
et al. (1990). Figure \ref{Pix-Vel} shows the expansion velocity
profiles for the NS matter as a function of mass coordinate
$m$. The numbers mark the time (in seconds)
from the loss of hydrodynamic stability by the star.
The expansion begins from the stellar surface and the
rarefaction wave reaches the center by $t_5=0.371$. By
this time almost the entire star already expands with
a speed $\sim 0.1$ of the light speed and the acceleration
wave travels in the opposite direction, outward. In
this case, an important event occurs: approximately
between $t_6=0.373$ and $t_7=0.375$ this wave crosses
the boundary of the NS mantle–crust at $m\approx 0.08M_\odot$.
While accelerating along a sharply falling density
profile (Fig. 1), it turns into a shock wave and heats
the matter.

\begin{figure}[ht]
	\begin{center}
	 \includegraphics[width=.95\textwidth]{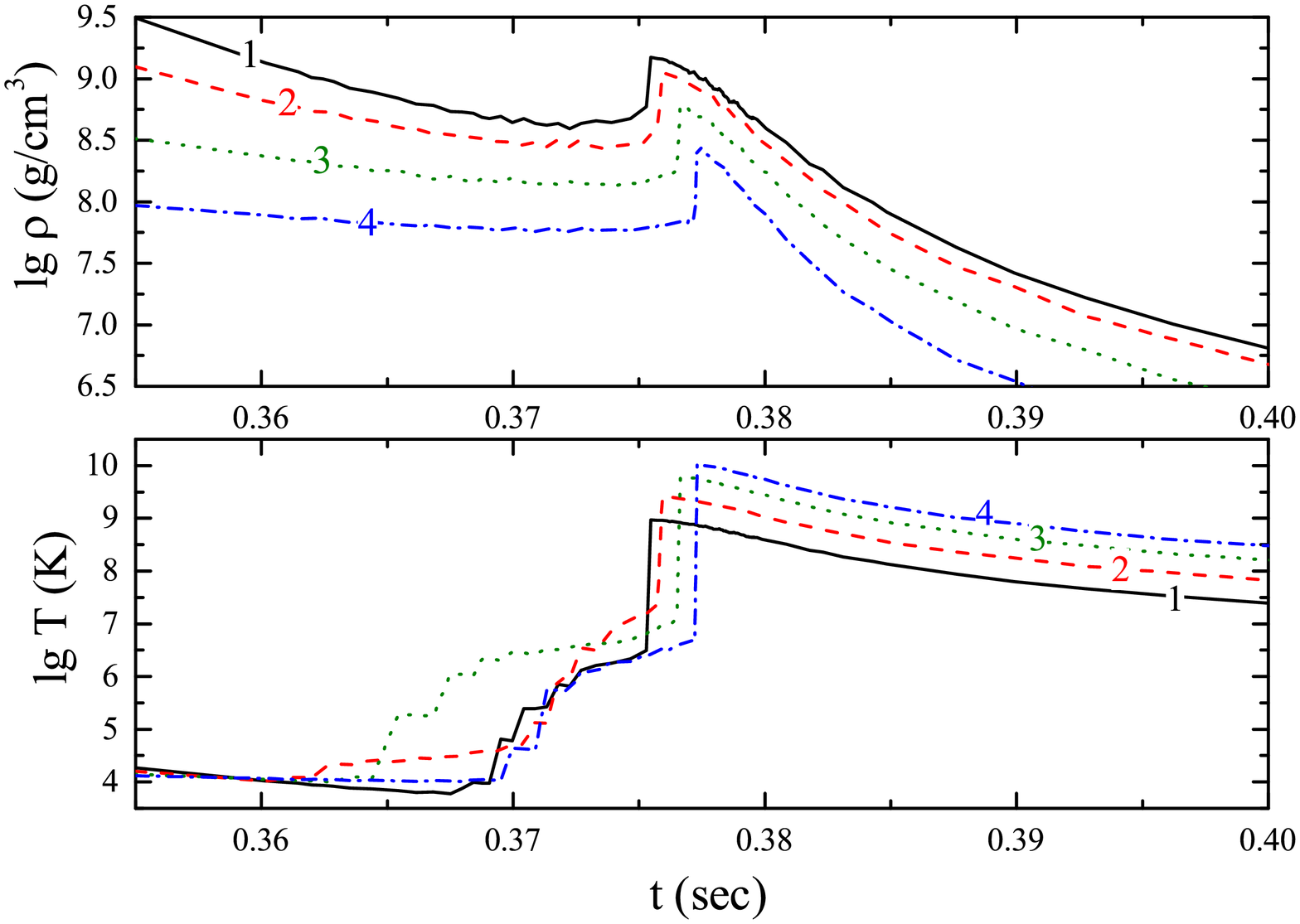}
\end{center}
	\caption{Evolution of the density and temperature along the four trajectories 1–4 considered as a function of time.} \label{Pix-Rho-T}
\end{figure}

The behavior of the density and temperature on
the four chosen trajectories (see Table 1 and Fig.~\ref{Pix-LMNS})
is shown in Fig.~\ref{Pix-Rho-T}. Initially, the density and temperature
drop due to the overall expansion. Subsequently,
at $t\approx 0.365$, the matter begins to be heated
by the acoustic oscillations inevitably generated as
the central part of the star expands. Propagating
away from the core along a falling density profile,
accelerating, and turning into weak shocks, these
small perturbations heat the matter. Later, already
by $t\approx 0.375$, a strong shock approaches the stellar
layers under consideration (its formation is seen in
Fig.~\ref{Pix-Vel} between the profiles marked by 7 and 8) and
causes a very fast increase in density (by a factor
of 2–3) and temperature (by orders of magnitude).
This time is especially difficult for the nucleosynthesis
calculations primarily because the thermonuclear
reaction rates change abruptly and the parameters of
the numerical algorithm are violated, which can lead
to a numerical instability of the algorithm for solving
the equations. After reaching the peak values, the
density and temperature continue to drop. As our
hydrodynamic calculations show, they decrease due
to the expansion of the matter in a regime faster than
its free expansion: $\rho\sim t^{-3.6}$, $T\sim\rho^{2/3}\sim t^{-2.4}$

It is easy to estimate the parameters of the NS
binary considered by us at the explosion time. The
Roche lobe radius $\WI{R}{R}$ for the low-mass NS approximately
coincides with its radius, i.e., $\WI{R}{R}\approx\WI{R}{s}\approx 270$~km (Fig.~\ref{Pix-LMNS}). The separation $a$ between the binary
components is related to the Roche lobe size by an
approximate relation (Paczynski 1971):
\begin{equation}
\frac{\WI{R}{R}}{a}\approx 0.462\left(\frac{q}{1+q}\right)^{1/3},
\end{equation}
where $q=m_2/m_1$ is the component mass ratio.
Taking $m_1=1.4M_\odot$ and $m_2=0.1M_\odot$ for our
estimate, we will get $a\approx 1441$~km. The escape
speed from the surface of the low-mass NS is $\WI{V}{esc}=\sqrt{\frac{2Gm_2}{\WI{R}{s}}}\approx 10^9~\mbox{cm}/\mbox{sec}$; the escape speed from the field
of the massive component is $\WI{V}{esc}=\sqrt{\frac{2G m_1}{a}}\approx 1.6\times 10^9~\mbox{cm}/\mbox{sec}$ Thus, the matter ejected during the low-mass
NS explosion has speeds (Fig.~\ref{Pix-Vel}) that exceed
the escape speed at least by several times. In reality,
in these simples estimates we disregarded the orbital
motion of the components, but detailed numerical
simulations (Manukovskiy 2010) of the expansion
of the matter in the NS binary under consideration
confirm our conclusion.

\section*{PRODUCTION OF HEAVY ELEMENTS
DURING A LOW-MASS NS EXPLOSION}
\noindent  We calculated the nucleosynthesis of heavy elements
along the above-described most typical trajectories
in the off-line mode. We determined the
evolution of the composition at a temperature $T_9>7$
in the NSE approximation (NSE stands for nuclear
statistical equilibrium) and used the SYNTHeR code
(Korneev and Panov 2011) to calculate the elemental
abundances as the temperature decreased. The
transition regime, when the over-compressed matter
at a subnuclear density whose composition is defined
by the equation of state rapidly passes to the qualitatively
different state of a dense hot plasma described,
in particular, by the Boltzmann–Maxwell equations
and the kinetic model of nucleosynthesis developed
for such conditions (Blinnikov and Panov 1996), is
most difficult for the si\-mu\-la\-ti\-ons. Therefore, the
transition from the subnuclear matter composed of
exotic neutron-overrich nuclei to the nuclear reaction
network was made formally. The basis formaking this
transition was the conservation of the preexplosion
local electron fraction $\WI{Y}{e}$ (which determines the degree
of matter neutronization) up to the onset of nucleosynthesis.
As the model is developed further, this
problem will be solved iteratively in the nucleosynthesis
calculations in combination with the expansion
hydrodynamics of the matter both by tracing its local
heating by the shock and by taking into account the
energy release during beta decay and fission.

\begin{figure}[ht]
 \vspace{1.cm}
 	\begin{center}
 	\includegraphics[width=.95\textwidth]{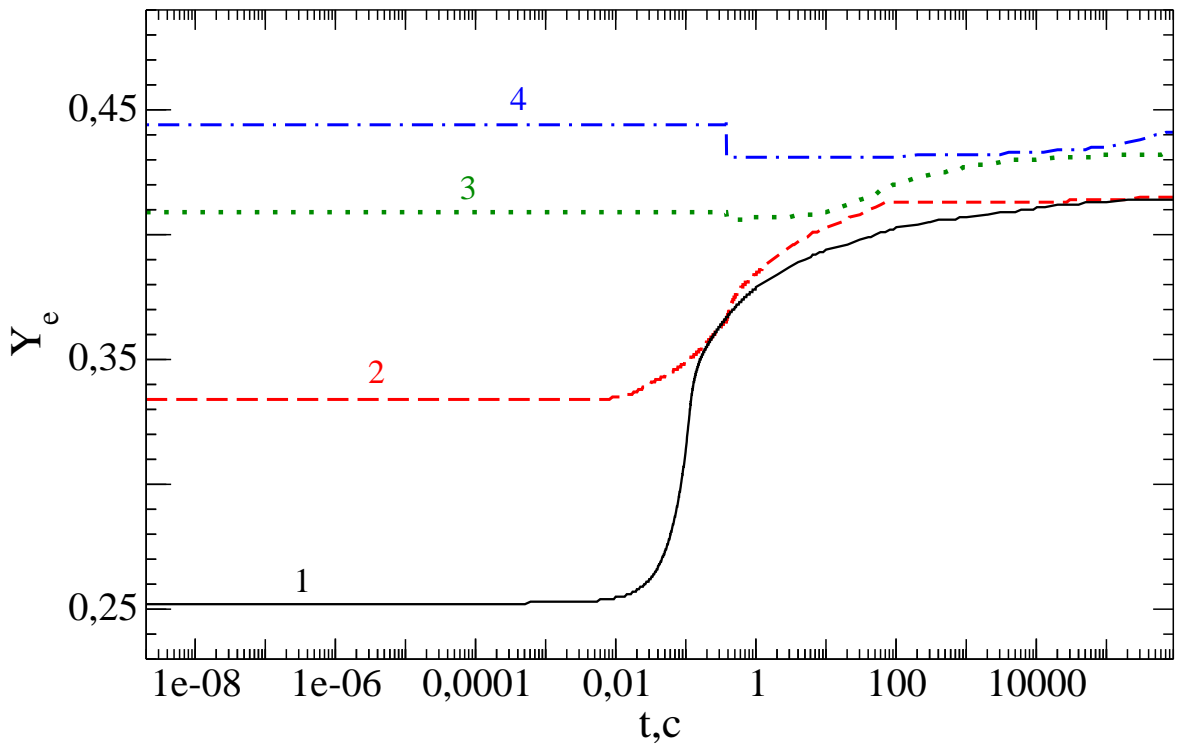}
    \end{center}
 	\caption{Electron fraction $\WI{Y}{e}$ versus time on the four chosen trajectories. The numbers near the curves are the trajectory
numbers.} \label{Fg4-Ye}
 \end{figure}

Below we discuss the results of our nucleosynthesis
calculations for four trajectories characterized
by different initial  $\WI{Y}{e}$  and initial isotopic compositions
(see Table 1). Figure~\ref{Fg4-Ye} shows the evolution of the
electron fraction, with the increase in $\WI{Y}{e}$ (at its initial
values less than 0.4) implicitly reflecting the intensity
and duration of the r-process begun after the second
expansion wave $t \approx 0.37$. The electron fraction $\WI{Y}{e}$
changes most dramatically due to the beta decays
both along the path of the r-process and at the cooling
stage as a result of the decay of unstable isotopes and
the production of stable elements only if $\WI{Y}{e} \le 0.4$. The
role of thermonuclear reactions in heating the matter
during the shock passage for variants 1 and 2 is
relatively minor (variant 2) or small (variant 1) and the
electron-to-proton ratio changes monotonically as
the beta decays of the neutron-rich nuclei produced
predominantly in the r-process occur. $\WI{Y}{e}$ in nucleosynthesis
increases along trajectory 1 considerably
longer than along trajectory 2 due to the beta decay
of a large number of long-lived isotopes of rare-earth
and trans-uranium elements. Under strong heating
(which is especially clearly seen for variant 4) the
role of thermonuclear reactions in the matter during
the shock passage increases, but the numerical effect
related to the rapidly varying processes at the shock
front can also have a noticeable influence on the jump
in $\WI{Y}{e}$. This aspect of the problem requires additional
studies.

\begin{figure}[ht]
\vspace{1.cm}
	\begin{center}
	\includegraphics[width=.95\textwidth]{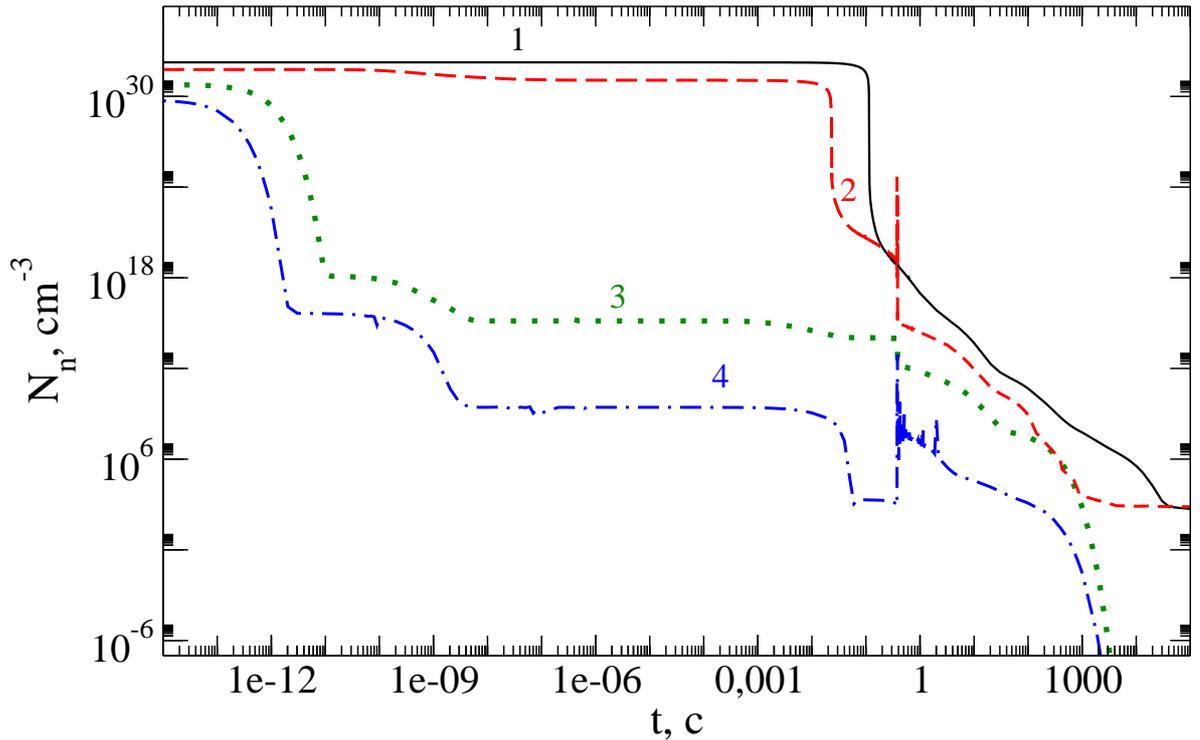}
  \end{center}
	\caption{Number density of free neutrons $N_n$ versus time on the four chosen trajectories. The numbers near the curves are the
trajectory numbers.} \label{Fg5-Nn}
\end{figure}

\begin{figure}[ht]
 	\begin{center}
 	\includegraphics[width=.95\textwidth]{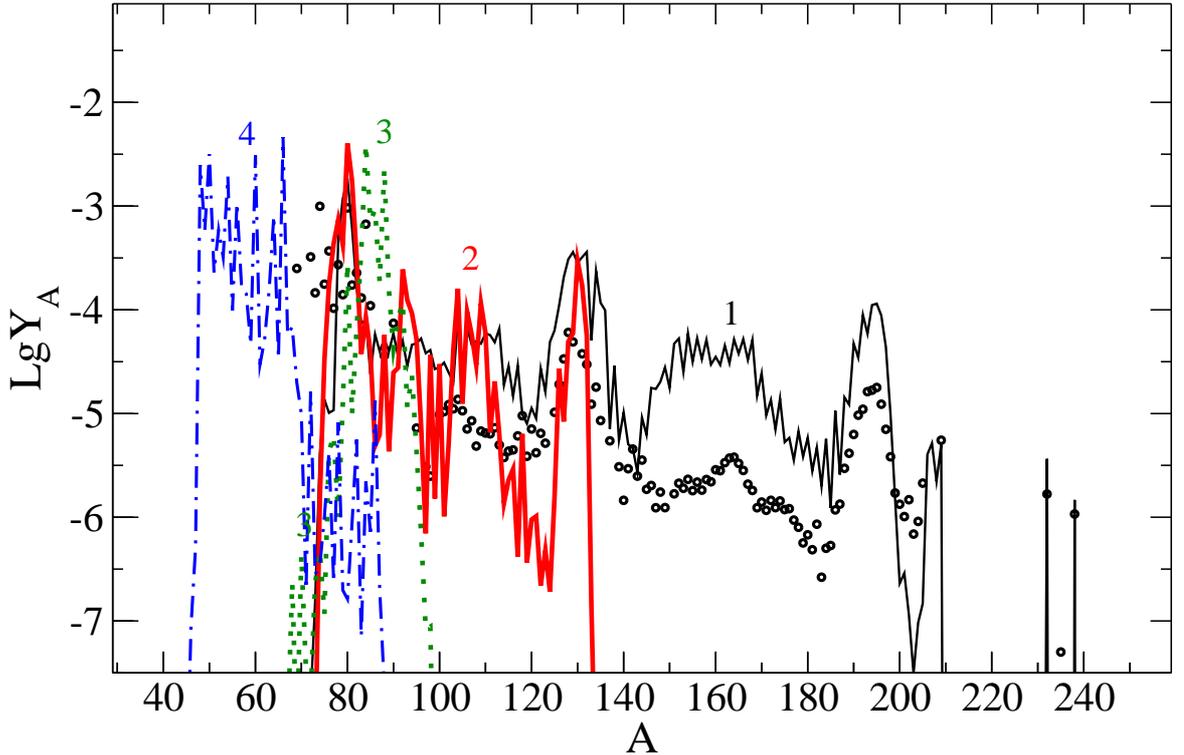}
   \end{center}
 	\caption{Abundances of the chemical elements produced at the end of nucleosynthesis along the four chosen trajectories. The
dots indicate the abundances of the heavy elements produced in the Solar System. The numbers near the curves are the
trajectory numbers.} \label{Fg6-YA}
 \end{figure}

Figure~\ref{Fg5-Nn} shows the evolution of the number density
of free neutrons as the r-process develops. It
follows from the figure that a level of free neutrons
$N_n \ge 10^{22}$~cm$^{-3}$ sufficient for the r-process to proceed
is maintained for several hundred milliseconds,
which is enough to produce all of the heavy nuclei up
to uranium only along trajectories 1 and 2 (Fig.~\ref{Fg6-YA}).
Accordingly, for variants 3 and 4, for which the initial
neutron excess decreases very rapidly during the
transition from a subnuclear density to densities of
the order of the hot-wind densities, the r-process does
not proceed and new elements are produced predominantly
through the $(\alpha,X)$--reactions, including those
produced during the matter heating by the shock and
the burst of nucleosynthesis due to the acceleration of
thermonuclear reactions.

The jumps in $\WI{N}{n}$ (variants 2 and 4) seen in Fig.~\ref{Fg5-Nn}
at $t\sim 0.37$ or the sharp absorption of neutrons (in
variant 3) occur with the switch-on of thermonuclear
reactions during the shock passage and are
largely attributable to the numerical effects whose
influence on the solution is minor. Note that the
significant and gradual decrease in $N_n$ at short times
$t<10^{-12},10^{-8},0.01$~s is determined mainly by the
switch-on of charge-exchange nuclear reactions.

The peaks in the region of atomic masses with
$A\sim 130$ and $196$ for variants 1 and 2 (Fig.~\ref{Fg6-YA}) are well-structured
and are consistent with the observations in
magnitude and place. The rare-earth peak $A\sim$ 160--170 for variant 1 is formed, which is even excessive
compared to the observational data.

\section*{CONCLUSIONS}
\noindent The evolution of close binaries can be different:
under some conditions a NS merger occurs; under
other ones a mass transfer between the companions
(stripping) is possible, which ends with a low-mass
NS explosion. The masses of the binary components
must differ greatly for the stripping mechanism to be
realized.

The fraction of binary stars with a low-mass companion
among the entire population of binary NSs
is apparently small. Such a binary configuration,
whose formation probability is yet to be determined,
will represent a fraction of the stripping mechanism
of GRBs in their total population.

 The original results obtained from our nucleosynthesis
calculations in the scenario for the evolution
of two NSs with significantly different masses show
that in the stripping scenario during the evolution
of two NSs part of the crust and mantle matter is
neutronized strongly enough for the r-process to proceed
in it during the explosion and expansion with the
production of a large amount of heavy elements. The
abundance curve of the heavy nuclei $Y(A)$ produced
during low-mass NS disruption, on the whole, agrees
well with both heavy-element abundance observations
and heavy-element abundance calculations for
a classical NS merger. For some trajectories the
heavy-element abundance combines the abundance
of the ``heavy'' fraction of the elemental abundance
typical for the ejection in the NS merger scenario and
the “light” component forming in the winds from a
hot massive neutron remnant in the same NS merger
scenario. It follows from Fig.~\ref{Fg6-YA} that when the third
peak is formed (variant 1), the ``iron'' peak of elements
with mass numbers $\sim80$ does not burn out, which
differs from the heavy-element production dynamics
in the NS merger scenario, where the second and
third peaks are formed (the main r-process) in the
jets or only the first and second peaks are formed (an
incomplete r-process in the wind).

The nucleosynthesis was calculated offline on fixed
tracks, without integrating the nucleosynthesis contribution
for all the possible trajectories, which allows
the difference between the nucleosynthesis calculations
for different layers of ejected matter to be estimated.
In fact, this is the first step in the project of
research on the stripping scenario and the nucleosynthesis
proceeding during the explosion of a low-mass
remnant. Our final goal is a self-consistent calculation
of the low-mass NS explosion that takes into account,
among other things, the additional heating of
the matter during intense nucleosynthesis due to beta
decay and fission. We are working on this problem at
present.

\section*{ACKNOWLEDGMENTS}
We are grateful to S.I. Blinnikov and D.K. Nadyozhin
for the discussion of physical processes at the final evolutionary
stages of a NS binary and the equation of state for
superdense matter and to N.I. Kramarev for his participation
in the discussion and his interest in the work. This work was supported by the Russian Foundation
for Basic Research (project no. 18{-}29{-}21019~mk).

\pagebreak

\end{document}